\documentclass[conference]{IEEEtran}
\usepackage[T1]{fontenc}
\usepackage[utf8]{inputenc}

\usepackage{amsmath}
\usepackage{amssymb}
\usepackage{amsfonts}
\usepackage{bm}                  % Bold math: \bm{x}

\usepackage{graphicx}
\graphicspath{{Figures/}}        % All \includegraphics paths relative to Figures/
\usepackage[dvipsnames]{xcolor}

\usepackage{booktabs}            % \toprule \midrule \bottomrule
\usepackage{multirow}
\usepackage{tabularx}
\usepackage{array}

\usepackage{algorithm}
\usepackage{algpseudocode}

\usepackage{cite}                % Compressed IEEE-style citations [1]-[3]
\usepackage[colorlinks=true,
            linkcolor=blue,
            citecolor=blue,
            urlcolor=blue]{hyperref}
\usepackage{cleveref}

\usepackage{balance}             % \balance on last page to equalize columns
\usepackage{microtype}           % Better line breaking / justification
\usepackage{url}

\title{Clockwork: Scheduling Compound AI Systems in Space Data Centers}

\author{
  \IEEEauthorblockN{Milos Gravara}
  \IEEEauthorblockA{
    Distributed Systems Group\\
    TU Wien\\
    m.gravara@dsg.tuwien.ac.at
  }\and
  \IEEEauthorblockN{Andrija Stanisic}
  \IEEEauthorblockA{
    Distributed Systems Group\\
    TU Wien\\
    a.stanisic@dsg.tuwien.ac.at
  }
  \and
  \IEEEauthorblockN{Stefan Nastic}
  \IEEEauthorblockA{
    Distributed Systems Group\\
    TU Wien\\
    s.nastic@dsg.tuwien.ac.at
  }
}

\begin{document}

\maketitle

% --- Abstract & Keywords ---
\begin{abstract}
    Space data centers built on LEO constellations create an opportunity to serve Compound AI workflows close to where observations are captured. Serving such a workflow for an area of interest (AoI) requires deciding which orbital planes to provision and where to place each workflow component, while orbital motion continuously changes which satellites can serve the AoI. Provisioning must therefore be planned as a sequence of decisions over time, and these decisions are coupled because changing component placements or activating new planes incurs costs. To address this, we present Clockwork, a framework for scheduling Compound AI workflows on space data centers under end-to-end latency SLOs. Clockwork has two components: i) Castor jointly selects active planes and component placements for each constellation snapshot, while ii) Pollux connects Castor's candidate provisioning states across predicted orbital evolution to construct a low-cost schedule over time. On our experimental two-shell constellation, Castor serves 81\% of instances at a 300~ms SLO, while Pollux reduces mean total scheduling cost by 56\% on average and by up to 63\% compared to heuristic baselines.
\end{abstract}


% --- Body sections (one file each) ---
% Sections/introduction.tex

\section{Introduction}
\label{sec:introduction}

The continued expansion of data centers is increasingly limited by power supply and cooling requirements, motivating the exploration of orbital infrastructure for large-scale computing~\cite{dclimits2026}. The orbital infrastructure already operates at scale, with Starlink alone operating more than seven thousand satellites and constellations such as Iridium and OneWeb providing global connectivity for years~\cite{ali2026starlink}. Computation is following connectivity, as Starcloud has launched H100-class GPUs, SpaceCloud processors fly on commercial missions, and SpaceX plans orbital data center satellites sustaining 120kW of compute each~\cite{starcloud2026, unibap2026}. A decade of research on orbital edge computing anticipated this trajectory~\cite{denby2020orbital, chen2026space}. Space data centers thus form an emerging tier of the computing continuum, and the question of how to manage their resources follows the hardware into orbit.

This shift is primarily driven by the computational demand of modern AI~\cite{aidemand2026}. The AI services heading to orbit rarely consist of a single model, as the AI landscape is evolving from monolithic models toward system-centric approaches that coordinate multiple specialized AI and non-AI components~\cite{gravara2026design}. Known as Compound AI, this approach combines specialized models and software components into workflows of components orchestrated to solve complex AI tasks~\cite{gravara2026compass}. Satellite platforms already host workloads of this shape, including remote sensing pipelines~\cite{jiang2025efficient}, wildfire monitoring workflows~\cite{pusztai2024hyperdrive}, and language model inference in space~\cite{wang2026spacemoe}, and the workload mix is expected to keep widening~\cite{chen2026space}.

A natural use case for this tier is provisioning constellation compute for an area of interest (AoI)~\cite{bhosale2024krios,zheng2025sdn}. A tenant names a region, supplies a Compound AI system such as wildfire monitoring, and specifies an end-to-end latency SLO. The platform must decide which orbital planes to activate, where to place each component, and when to change these decisions. These decisions determine whether the workflow meets its SLO and what the service costs over time.

However, to make these decisions, the platform must address two challenges. First, constellation hardware can be heterogeneous, so the resources required by different Compound AI components may be located on different satellites and orbital planes~\cite{10.1007/978-3-032-35251-4_24}. A workflow may therefore need to span multiple satellites or planes, which introduces additional ISL communication and requires provisioning compute resources across those locations.

Second, a placement that satisfies the latency SLO at one point in time may no longer satisfy it as the constellation moves relative to the AoI~\cite{pusztai2024hyperdrive, wang2026spacemoe}. Changes in satellite visibility and propagation latency alter the communication paths available to the workflow over time. The platform may need to place components on different satellites and orbital planes to restore SLO compliance, incurring additional provisioning costs. Therefore, the platform must decide not only where the workflow should run, but also when changing that placement is worth its cost.

Existing work on LEO constellations largely decides where a workload runs while treating the available satellites as fixed. Workflow schedulers place multi component pipelines on satellites under latency objectives~\cite{pusztai2024hyperdrive,li2025orbitchain}. Methods based on time expanded LEO graphs limit migrations of service chains, typically for uniform network functions on individual satellites~\cite{mahyoub2026scalable,liu2026vmr,liu2025switching}. Region bound orchestration maintains application reachability over a ground zone through predictive network handovers without accounting for the cost of resources that remain provisioned or change~\cite{bhosale2024krios,zheng2025sdn}. On the ground, provisioned compute resources remain fixed relative to the demand they serve. In orbit, the planes used for compute and access move relative to the region, so the platform must change provisioning states over time. This creates a temporal provisioning problem that remains unaddressed by existing approaches, because orbital motion couples which resources are used, where workflow components are placed, and when the provisioning should change.

In this paper, we introduce Clockwork\footnote{The name reflects the predictability of orbital motion, which enables the framework to plan provisioning decisions in advance. It is unrelated to the Clockwork inference-serving system of Gujarati et al.~\cite{gujarati2020clockwork}.}, a novel framework for provisioning Compound AI systems in space data centers under an end-to-end latency SLO. Clockwork uses predicted orbital geometry to plan provisioning decisions across constellation snapshots. To do so, Clockwork relies on two novel algorithms. Castor constructs candidate provisioning states for individual snapshots, and Pollux connects them across snapshots while accounting for provisioning and reprovisioning cost. The resulting schedule specifies the shell, active planes, and component placements used over time.

The main contributions of our work include:

\begin{itemize}
\item \textbf{Clockwork}: a novel planning framework and system model for scheduling Compound AI systems in space data centers. Clockwork represents a provisioning state through its shell, component placements, and active planes, while allowing ingress and egress to change with orbital motion without reprovisioning. It connects snapshot feasibility under the latency SLO with provisioning and reprovisioning costs over time, providing a common formulation for placement and scheduling decisions.

\item \textbf{Castor}: a per-snapshot provisioning algorithm that jointly selects component placements and active planes under the latency SLO. Castor searches a layered graph using label dominance and latency and cost bounds, returning a minimum-cost feasible state over candidate satellites and alternatives within a specified cost-rate margin. At various SLOs in the reference simulation, Castor finds minimum-cost feasible states in up to 98\% of instances, representing an increase of up to 40\% in the fraction of feasible instances compared with the heuristic baselines.

\item \textbf{Pollux}: a cyclic scheduling algorithm that connects Castor's candidate states across predicted constellation snapshots. Pollux reevaluates their feasibility over time and uses dynamic programming to account for provisioning and reprovisioning costs when deciding whether to retain or change a state. In the default configuration, Pollux reduces mean total scheduling cost by 63\% relative to baselines while preserving the same SLO attainment.
\end{itemize}

% The rest of this paper is organized as follows. \Cref{sec:motivation} presents the motivating use case, \Cref{sec:framework} the Clockwork framework and system model, and \Cref{sec:castor,sec:pollux} the Castor and Pollux algorithms. \Cref{sec:evaluation} evaluates Clockwork, \Cref{sec:related} reviews related work, and \Cref{sec:conclusion} concludes.
\section{Motivational Use Case and Background}
\label{sec:motivation}

% Between 2014 and 2023, wildfires caused an estimated \$ 106 billion in global economic losses, while early detection gives responders more time to contain a fire before it spreads~\cite{united2025global}. 

% Between 2014 and 2023, wildfires caused an estimated \$~106 billion in global economic losses~\cite{united2025global}. Early detection can limit this damage by giving responders more time to contain fires before they spread~\cite{united2025global}. 

In remote regions, early wildfire detection is difficult because these regions often lack sufficient sensing and connection to the terrestrial communication infrastructure. Low Earth orbit (LEO) satellites can address this coverage gap by repeatedly imaging the affected region~\cite{10.1007/978-3-032-35251-4_24}. However, transmitting every captured image for processing on the ground is impractical because satellites can collect more imagery than their intermittent ground links can transmit~\cite{denby2020orbital}. Instead, AI models can process observations on the satellites and return compact alerts rather than complete images~\cite{moussa2025pyrofocus}. Such processing can combine object detection to identify fire or smoke, image classification to examine the detected region, and report generation to produce an alert for responders. A pipeline that coordinates these specialized models is a Compound AI system~\cite{gravara2026compass, pusztai2024hyperdrive}. Running such a workflow in orbit requires placing its components on compute resources distributed across the LEO constellation.

\begin{figure}[t]
  \centering
  \includegraphics[width=0.8\columnwidth]{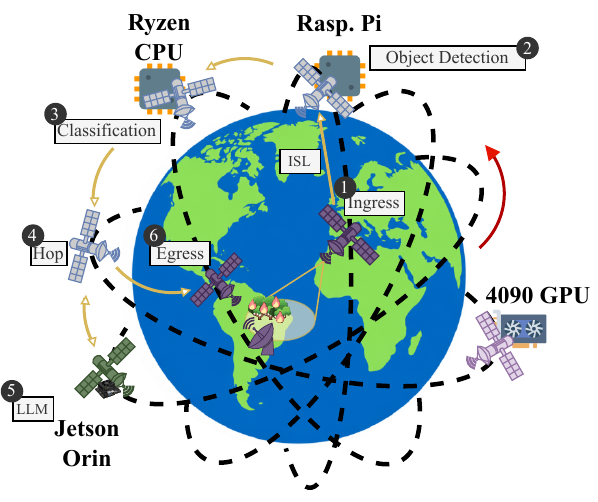}
  \caption{Wildfire detection use case}
  \label{fig:wildfire}
\end{figure}

% Meanwhile, the compute substrate available in orbit is expanding beyond individual satellite processors. Server-class accelerators have already flown in space, and proposed space data-center constellations envision substantially larger compute deployments~\cite{starcloud2026,fcc2026odc,unibap2026}. These systems inherit the organization of large LEO constellations such as Starlink~\cite{ali2026starlink}. 

Satellites that share an altitude and inclination form a shell, each shell is divided into orbital planes, and each plane is a ring of satellites on the same orbit~\cite{walker1970some}. Inter-satellite links (ISLs) connect a satellite to its neighbors in the same plane and in the adjacent planes, so the workflow components are not confined to the satellite that captured the image but can be placed across satellites in different planes of a shell~\cite{wood2001effects,ali2026starlink}. The available hardware is also far from uniform. Radiation-hardened processors such as the RAD5545 deliver a few GFLOPS, embedded accelerators of the Jetson class fly on commercial missions, and a single H100-class GPU has reached orbit, so a constellation holds many CPU satellites, fewer with small accelerators, and only a handful with large ones, concentrated in a few planes~\cite{bae2023rad5545,unibap2026,starcloud2026}. Unlike a terrestrial cluster, however, the entire substrate moves continuously relative to the Earth's surface, changing which satellites and planes are positioned to serve a particular region over time~\cite{ali2026starlink}. A LEO satellite travels at about 7.6\,km/s and completes an orbit in roughly 95 minutes, so it stays above a burning region for only a few minutes before the next satellite, often in another plane, takes its place. A wildfire alert must therefore be executed on a substrate whose usable resources change every few minutes.

% Consider a provider that offers a constellation as a service. A tenant names an area of interest (AoI), a compound AI workflow, and an end-to-end latency SLO. The provider serves the request by activating orbital planes and placing the workflow components on their satellites, and charges it for each active plane and each satellite that computes.  

% \Cref{fig:wildfire} illustrates this service for a wildfire request at one snapshot of the constellation.

\Cref{fig:wildfire} illustrates how the Compound AI system for wildfire detection can be provisioned at one snapshot of the constellation. In this example, to serve the request, the platform needs an active plane with a satellite above the fire, since that satellite captures the image and is the ingress of the request (1). It places the detector (2) and the classifier (3) on a CPU satellite of the same plane, since neither requires an accelerator. For the report generator it needs an accelerator, and the nearest one is located in a neighboring plane, so it activates that plane as well and forwards the classified result over ISLs (4) to generate the alert (5). It cannot use the large accelerator, although that one is faster, because its plane is not in view of the AoI and the additional ISL distance exceeds the SLO. Finally, it needs a satellite in view of the AoI in an active plane to return the alert, the egress of the request (6), which is not necessarily the ingress satellite. 

Even at one constellation snapshot, determining how to provision the workflow is nontrivial. Several satellites may provide ingress and egress, while compatible compute resources can lie in different planes. Activating additional planes exposes more hardware and can reduce execution or communication latency, but increases provisioning cost. Component placements also interact through hardware compatibility, satellite capacity, payload transfers, and the end to end SLO. The platform must therefore jointly select the active planes and component placements, since locally attractive choices can make the complete workflow too expensive or infeasible. 

Solving this placement problem independently at each snapshot is still insufficient. As satellites move relative to the AoI, visibility and propagation latency change, so the cheapest feasible provisioning state can change within minutes. Following each change can repeatedly activate planes or move components and incur reprovisioning cost, while retaining a slightly more expensive state can avoid later transitions. Orbital motion makes these future changes predictable, but knowing them does not determine which states should be retained or replaced. The platform must therefore plan a sequence of provisioning states over the predicted horizon, balancing the cost of each state against the cost of changing it while maintaining the SLO whenever service is feasible.

% The arrangement in the figure remains valid only while its satellites are in view, which at LEO speeds is a matter of minutes, so the provider has to find hundreds of such arrangements for one request over a single day. Each change can require activating new planes and starting components on new satellites before the previous ones are released, so a transition can cost the tenant both arrangements at once and can miss the SLO until the new components are running. Changing at every opportunity multiplies these transition costs, while keeping an arrangement after its satellites leave view violates the SLO. Provisioning is therefore a decision in time as well as in space. The provider must select an arrangement at each snapshot and decide when a change is worth its cost. Orbital motion, however, is deterministic, and the position of every satellite relative to the AoI is known in advance for any planning horizon. This predictability allows the provider to compute the complete sequence of arrangements before it is needed. Serving compound AI workflows on a LEO constellation thus becomes a planning problem of which planes to activate, where to place each component, and when to change, minimizing cost while satisfying the SLO at each serviceable planning epoch.

\section{Clockwork Framework}
  \label{sec:framework}

\begin{figure*}[t]
\centering
\includegraphics[width=0.95\textwidth]{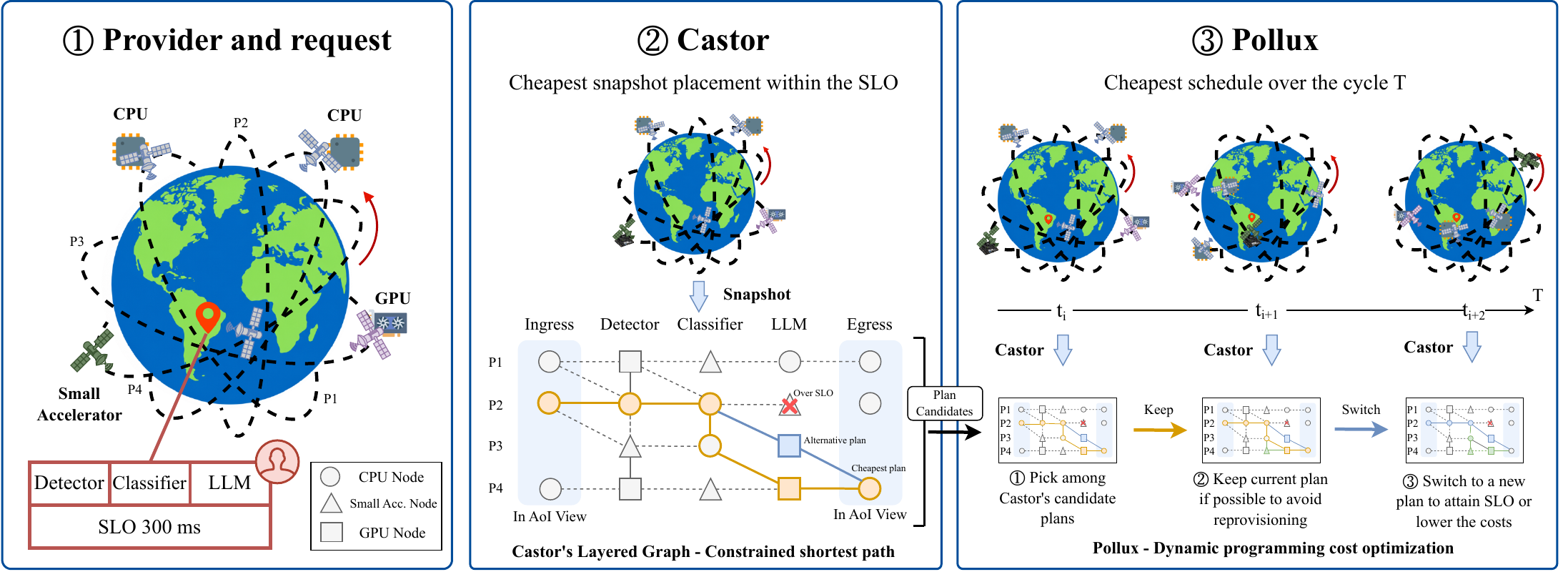}
\caption{Clockwork overview. (1) The provider offers shells of orbital planes with heterogeneous hardware; a request names an AoI, a Compound AI system workflow, and a latency SLO. (2) Per snapshot, Castor searches a
layered graph of candidate satellites for a minimum-cost feasible state and retains near-minimum-cost alternatives. (3) Per cycle, Pollux selects a state sequence from the resulting pool, using future snapshots to minimize necessary reprovisionings and cost.}

\label{fig:clockwork-overview}
\end{figure*}

Clockwork plans where a Compound AI workflow is provisioned on an orbital compute infrastructure and when that provisioning changes over time with respect to an AoI. Its inputs are a provider description of the constellation, hardware, network, and provisioning costs, together with a service request containing the workflow, area of interest, and latency SLO. The provider also supplies execution profiles for each workflow component on each supported hardware class. Clockwork produces a provisioning schedule that specifies the shell, active planes, and component placements used over time.

Clockwork separates this problem into two stages. Castor determines feasible provisioning states for individual constellation snapshots and identifies low cost alternatives. Pollux then connects these states across snapshots and selects when to retain or change them according to provisioning and reprovisioning cost. Ingress and egress satellites may change while a provisioning state remains fixed, allowing the same state to remain useful as satellites move relative to the AoI.

Clockwork assumes stateless workflow executions with no mutable application state shared across requests. The provider prepares the software and model artifacts required by a scheduled provisioning state. Clockwork plans provisioning and accounts for the cost of changing it. Deployment execution and application state management are outside its scope.

\subsection{Orbital Substrate and Provisioning Request}
\label{sec:framework:infrastructure}

The provider operates a set of Walker shells $\mathcal{K}$. Each shell $k \in \mathcal{K}$ contains satellites $\mathcal{S}_k$ organized into orbital planes $\mathcal{P}_k$. Each satellite $s$ belongs to plane $p(s)$, has hardware class $\tau_s$, and provides compute capacity $\kappa_s$.

Satellites communicate through ISLs connecting neighbors within the same plane and in adjacent planes. Walker Star shells form a cylindrical topology, while Walker Delta shells also connect the first and last planes and form a toroidal topology~\cite{wood2001effects}. Connections remain fixed in the model, while propagation latency changes with orbital motion. We assume fixed transmission rates and omit signal degradation and link failures. We denote the propagation latency of the minimum-propagation ISL path between satellites $s$ and $s'$ at time $t$ by $d_t(s,s')$. Shells are not connected by ISLs, so each provisioning state uses one shell. Activating a plane allows its satellites to host workflow components and serve as ingress or egress. Satellites in inactive planes may still forward ISL traffic.

A service request specifies an area of interest $A$, a fixed Compound AI workflow with components $\mathcal{C}=(c_1,\ldots,c_m)$, and a latency SLO $\delta$. Each component $c_i$ has compute capacity demand $\rho_i$ and execution latency $\ell_i(\tau)$ on hardware class $\tau$. We set $\ell_i(\tau)=\infty$ when an execution profile is unavailable, which denotes incompatible hardware. The input size is $b_0$, and $b_i$ denotes the output size of component $c_i$ in bits.

We represent the AoI by a fixed ground point that also locates the receiving ground station. At time $t$, a satellite is visible from the AoI when its elevation meets or exceeds the configured minimum elevation. We denote the visible satellites of shell $k$ by $\mathcal{V}_k(A,t)$. An ingress satellite captures the input over the AoI, and an egress satellite returns the workflow result to the ground station. The two satellites may differ. The propagation latency between the AoI and satellite $s$ is denoted by $\alpha_t(A,s)$.

\subsection{Provisioning State and Feasibility}
\label{sec:framework:state}

A provisioning state specifies a shell, a placement of the workflow components, and the active orbital planes:
\begin{equation}
    q = (k,\pi,\mathcal{Y}),
    \qquad
    \pi:\mathcal{C}\rightarrow\mathcal{S}_k,
    \qquad
    \mathcal{Y}\subseteq\mathcal{P}_k.
    \label{eq:provisioning-state}
\end{equation}

The mapping $\pi$ assigns each component to a satellite. Ingress and egress selections are not part of $q$ and may change without reprovisioning while suitable visible satellites remain available in the active planes.

Let $\mathcal{U}(q)$ contain the satellites hosting at least one workflow component. Each active plane $p$ incurs provisioning rate $\phi_p$, and each satellite in $\mathcal{U}(q)$ incurs rate $\chi(\tau_s)$ according to its hardware class. A satellite is charged once even when it hosts multiple components. The provisioning cost rate is
\begin{equation}
    C(q)
    =
    \sum_{p \in \mathcal{Y}} \phi_p
    +
    \sum_{s \in \mathcal{U}(q)} \chi(\tau_s).
    \label{eq:provisioning-cost}
\end{equation}

Let $s_i=\pi(c_i)$ denote the satellite hosting component $c_i$, and let $s_0$ and $s_{m+1}$ denote the ingress and egress satellites. A valid placement requires components to use compatible hardware in active planes and colocated components to fit within satellite capacity. At snapshot $t$, ingress and egress must also be visible from the AoI and belong to active planes:
\begin{equation}
\begin{aligned}
    p(s_i) &\in \mathcal{Y}
    && \forall i=1,\ldots,m, \\
    \ell_i(\tau_{s_i}) &< \infty
    && \forall i=1,\ldots,m, \\
    \sum_{i:s_i=s}\rho_i &\leq \kappa_s
    && \forall s \in \mathcal{S}_k, \\
    s_0,s_{m+1} &\in \mathcal{V}_k(A,t), \\
    p(s_0),p(s_{m+1}) &\in \mathcal{Y}.
\end{aligned}
\label{eq:placement-constraints}
\end{equation}

Let $\mathcal{Q}$ contain the provisioning states satisfying component placement, active-plane membership, hardware compatibility, and capacity constraints, independently of time-dependent access feasibility.

Clockwork assumes uncongested ISLs with transmission rate $r_{\mathrm{ISL}}$ and pipelined forwarding. For $i=0,\ldots,m-1$, $T_{i,t}(s_i,s_{i+1})$ denotes the transfer from the ingress or component $i$ to the next component. It is zero when $s_i=s_{i+1}$. Otherwise, it equals
\[
    d_t(s_i,s_{i+1}) + \frac{b_i}{r_{\mathrm{ISL}}}.
\]
Each component begins execution after receiving its complete input.

The final result may traverse ISLs before reaching the egress. The egress streams the result to the ground station as it arrives. We assume downlink rate $r_{\mathrm{down}}\leq r_{\mathrm{ISL}}$, so the final payload is serialized once at the downlink rate. The end-to-end latency from input capture to result delivery is
\begin{equation}
\begin{aligned}
    L_t(q,s_0,s_{m+1})
    ={}&
    \sum_{i=1}^{m} \ell_i(\tau_{s_i})
    +
    \sum_{i=0}^{m-1} T_{i,t}(s_i,s_{i+1}) \\
    &+
    d_t(s_m,s_{m+1})
    +
    \frac{b_m}{r_{\mathrm{down}}}
    +
    \alpha_t(A,s_{m+1}).
\end{aligned}
\label{eq:end-to-end-latency}
\end{equation}

Let $\mathcal{Q}_t \subseteq \mathcal{Q}$ denote the states for which some visible ingress and egress pair in the active planes satisfies the access constraints in~\Cref{eq:placement-constraints} and has $L_t(q,s_0,s_{m+1})\leq\delta$. The per-snapshot provisioning problem selects a feasible state with minimum provisioning cost rate:
\begin{equation}
    q_t^\star
    \in
    \arg\min_{q \in \mathcal{Q}_t} C(q).
    \label{eq:snapshot-objective}
\end{equation}

Castor searches this objective over its candidate satellites and returns low-cost feasible states for Pollux.

\subsection{Cyclic Provisioning Schedule}
\label{sec:framework:schedule}

Clockwork plans over snapshots $0=t_1<\cdots<t_N<T$, where $T$ is a common repeat period of the modeled shells. We set $t_{N+1}=T$ and $\Delta_j=t_{j+1}-t_j$. When the AoI-relative geometry repeats after $T$, the resulting provisioning schedule can be reused. Otherwise, Clockwork constructs a new schedule from updated orbital predictions.

A cyclic provisioning schedule is $\Gamma=(q_1,\ldots,q_N)$, where state $q_j$ applies during $[t_j,t_{j+1})$. Feasibility is evaluated at snapshot times and is not guaranteed throughout the intervals between them. Whenever $\mathcal{Q}_{t_j}$ is nonempty, the schedule must select a feasible state. When no feasible state exists, the provider may retain a state from $\mathcal{Q}$ or release all resources and enter the empty state $\varnothing$, with $C(\varnothing)=0$. Retaining resources through a service gap can avoid reprovisioning when service becomes possible again.

A transition to a nonempty state incurs reprovisioning cost $\lambda$ when it changes the component placement or activates a plane that was previously inactive. Deactivating planes is free, including releasing all resources. Let $R(q,q')$ equal one when the transition from $q$ to $q'$ incurs reprovisioning cost and zero otherwise. The transition from the final state $q_N$ back to $q_1$ follows the same rule.

The cyclic scheduling problem minimizes provisioning and reprovisioning cost over the repeat cycle:
\begin{equation}
\begin{aligned}
    \Gamma^\star \in \arg\min_{\Gamma}\quad
    &
    \sum_{j=1}^{N} C(q_j)\Delta_j
    +
    \lambda \sum_{j=1}^{N} R(q_j,q_{j+1})
    \\
    \text{s.t.}\quad
    &
    q_j \in \mathcal{Q}_{t_j},
    && \mathcal{Q}_{t_j} \neq \varnothing,
    \\
    &
    q_j \in \mathcal{Q}\cup\{\varnothing\},
    && \mathcal{Q}_{t_j} = \varnothing,
    \\
    &
    q_{N+1} = q_1.
\end{aligned}
\label{eq:cyclic-objective}
\end{equation}

Pollux constructs a cyclic schedule for this objective from the candidate states generated by Castor across snapshots.

\section{Castor: Per-Snapshot Provisioning}
\label{sec:castor}

Castor jointly selects component placements and active planes for a constellation snapshot to solve the restricted provisioning problem in~\cref{eq:snapshot-objective}. It returns a state pool $\mathcal{Q}_t^\eta$ containing a minimum-cost feasible state over its candidate satellites and nondominated alternatives within a cost-rate margin $\eta$. These alternatives give Pollux choices beyond the minimum-cost state at each snapshot.

\paragraph{Search representation}
Castor represents provisioning choices as a layered graph with an ingress layer, one layer per compound AI workflow component, and an egress layer. Each complete path selects ingress, component placements, and egress within one shell. To restrict the search, Castor includes all visible satellites and all accelerator-equipped satellites. For each of these, it also includes the $K$ nearest CPU satellites in the same shell and the $K$ nearest CPU satellites in the same plane, ranked by the propagation latency $d_t(s,s')$ defined in~\cref{sec:framework:infrastructure}.

A label represents a partial path through the graph:
\begin{equation}
    \sigma =
    (i, s, c_\sigma, l_\sigma, \mathcal{Y}_\sigma, \mathbf{u}_\sigma),
    \label{eq:castor-label}
\end{equation}
where $i$ and $s$ identify its current layer and satellite, $c_\sigma$ and $l_\sigma$ are its accumulated cost rate and latency, $\mathcal{Y}_\sigma$ contains its active planes, and $\mathbf{u}_\sigma$ records its per-satellite capacity usage. \Cref{alg:castor} searches the graph by advancing these labels through successive layers.

\paragraph{Building a plan}
Castor starts from each visible ingress and charges the activation of its plane (line~\ref{alg:castor:initialize}). It then places successive compound AI workflow components on satellites with compatible hardware and sufficient remaining capacity, following the constraints in~\cref{eq:placement-constraints} (line~\ref{alg:castor:extend}). Each placement adds the corresponding transfer and execution latency and charges newly used satellites and planes according to~\cref{eq:provisioning-cost,eq:end-to-end-latency}. Selecting a visible egress completes the path.

\paragraph{Pruning}
Castor estimates the minimum additional latency and cost needed to complete each partial path. The resulting lower bounds $\underline{L}(\sigma)$ and $\underline{C}(\sigma)$ include the accumulated latency and cost rate. Castor discards a path when these bounds rule out satisfying the SLO or remaining within the permitted cost margin (line~\ref{alg:castor:bounds}).

Castor also compares labels at the same satellite and layer using dominance~\cite{irnich2005shortest}. One label dominates another when it has no greater cost rate or latency, no greater per-satellite capacity usage, and an active-plane set containing the other's, with at least one strict relation. Castor discards the dominated label.

\paragraph{Output}
Castor retains complete paths within $\eta$ of the minimum cost rate found (line~\ref{alg:castor:margin}). It reconstructs their provisioning states and checks their costs and feasibility using the plan evaluator before returning $\mathcal{Q}_t^\eta$ (line~\ref{alg:castor:verify}). An empty pool means that no feasible state was found within the candidate satellites.

\begin{algorithm}[t]
\caption{Castor per-snapshot provisioning}
\label{alg:castor}
\begin{algorithmic}[1]
\scriptsize
    \Require Snapshot time $t$, request $\mathcal{R}$, margin $\eta$
    \Ensure State pool $\mathcal{Q}_t^\eta$

    \State $\mathcal{L}_0 \gets \Call{InitializeIngress}{t}$
        \label{alg:castor:initialize}
    \State $\overline{C} \gets \Call{InitialCostBound}{t}$
        \label{alg:castor:incumbent}

    \For{$i = 1,\ldots,m+1$}
        \State $\mathcal{L}_i \gets \varnothing$
        \ForAll{$\sigma \in \mathcal{L}_{i-1}$ and feasible successors $v$}
            \State $\sigma' \gets \Call{Extend}{\sigma,v,t}$
                \label{alg:castor:extend}
            \If{$\underline{L}(\sigma') \leq \delta$
                and $\underline{C}(\sigma') \leq \overline{C} + \eta$}
                \label{alg:castor:bounds}
                \State $\Call{ParetoInsert}{\mathcal{L}_i,\sigma'}$
                    \label{alg:castor:pareto}
            \EndIf
        \EndFor
    \EndFor

    \State $\mathcal{F} \gets \mathcal{L}_{m+1}$
        \label{alg:castor:final-labels}

    \If{$\mathcal{F} = \varnothing$}
        \State \Return $\varnothing$
    \EndIf

    \State $C^\star \gets \min_{\sigma \in \mathcal{F}} c_\sigma$
    \State $\mathcal{F}^{\eta} \gets
        \{\sigma \in \mathcal{F} \mid c_\sigma \leq C^\star + \eta\}$
        \label{alg:castor:margin}

    \State \Return
        $\Call{MaterializeAndVerify}{\mathcal{F}^{\eta},t}$
        \label{alg:castor:verify}
\end{algorithmic}
\end{algorithm}

\section{Pollux: Cyclic Schedule Construction}
\label{sec:pollux}

Pollux uses Castor to generate candidate provisioning states and connects them across constellation snapshots into a cyclic provisioning schedule $\Gamma$. Selecting the minimum-cost state at each snapshot can require frequent reprovisioning. Pollux instead accounts for both provisioning and reprovisioning costs in~\cref{eq:cyclic-objective}, allowing it to retain a more expensive state when doing so reduces total scheduling cost.

\paragraph{Candidate pool}
Pollux calls Castor at each snapshot and combines its returned state pools into $\mathcal{P}$ (line~\ref{alg:pollux:castor}). For each state, it also includes a state with the same component placement and only the active planes hosting components, allowing unused access planes to be released (line~\ref{alg:pollux:release}).

Pollux evaluates every candidate at every snapshot using the visible ingress and egress pairs in its active planes. The resulting table $\mathbf{F}$ records the lowest-latency pair and whether it satisfies the SLO according to~\cref{eq:end-to-end-latency} (line~\ref{alg:pollux:evaluate}). Thus, a state can remain feasible across snapshots even when its ingress and egress satellites change.

\paragraph{Permitted states}
At each snapshot, Pollux selects a feasible candidate whenever one exists. Otherwise, it may retain provisioned resources or release them by selecting the empty state. The permitted set at snapshot time $t_j$ is
\begin{equation}
    \mathcal{P}_j =
    \begin{cases}
        \mathcal{P} \cap \mathcal{Q}_{t_j},
            & \mathcal{P} \cap \mathcal{Q}_{t_j} \neq \varnothing, \\
        \mathcal{P} \cup \{\varnothing\},
            & \text{otherwise}.
    \end{cases}
    \label{eq:pollux-permitted}
\end{equation}
Pollux constructs these sets from $\mathbf{F}$ (line~\ref{alg:pollux:permitted}). Holding resources through a service gap can avoid reprovisioning when service becomes possible again. An empty feasible candidate set does not establish infeasibility over the full state space $\mathcal{Q}$.

\paragraph{Constructing the schedule}

Pollux uses dynamic programming to select a state sequence from the permitted sets (\cref{alg:pollux}). Because the schedule is cyclic, it can be cut at any snapshot without changing its cost. Pollux cuts the cycle at the snapshot with the fewest permitted states, rotates the sequence so that this snapshot is first, and considers each state in the resulting $\mathcal{P}_1$ as the initial state (lines~\ref{alg:pollux:cut}--\ref{alg:pollux:initialize}).

For each state, Pollux compares the cheapest predecessor requiring no reprovisioning with the cheapest predecessor overall plus reprovisioning cost $\lambda$, and selects the cheaper option after adding the provisioning cost for the current interval (lines~\ref{alg:pollux:global}--\ref{alg:pollux:update}). If no predecessor permits a free transition, that option has infinite cost. At the final snapshot, Pollux accounts for the transition back to the initial state and retains the minimum-cost cyclic sequence across all initial states (line~\ref{alg:pollux:backtrack}).

\paragraph{Output}

Pollux reconstructs the selected cyclic sequence by following the recorded predecessors (line~\ref{alg:pollux:backtrack}). Consecutive intervals assigned the same provisioning state form one schedule segment. The table $\mathbf{F}$ supplies ingress and egress selections at each snapshot, which may change within a segment. The returned sequence minimizes the cyclic objective in~\cref{eq:cyclic-objective} over Pollux's candidate pool.

\begin{algorithm}[t]
\caption{Pollux cyclic schedule construction}
\label{alg:pollux}
\begin{algorithmic}[1]
\scriptsize
    \Require Snapshot times $t_1,\ldots,t_N$, request $\mathcal{R}$,
        margin $\eta$, reprovisioning cost $\lambda$
    \Ensure Cyclic provisioning schedule $\Gamma$

    \State $\mathcal{P} \gets \bigcup_{j=1}^{N} \Call{Castor}{t_j,\mathcal{R},\eta}$
        \label{alg:pollux:castor}
    \State $\mathcal{P} \gets \Call{AddReleaseStates}{\mathcal{P}}$
        \label{alg:pollux:release}
    \State $\mathbf{F} \gets
        \Call{EvaluateStates}
        {\mathcal{P},(t_1,\ldots,t_N),\mathcal{R}}$
        \label{alg:pollux:evaluate}
    \State Construct $\mathcal{P}_1,\ldots,\mathcal{P}_N$
        from $\mathbf{F}$ using~\cref{eq:pollux-permitted}
        \label{alg:pollux:permitted}
    \State $k \gets \arg\min_{j} |\mathcal{P}_j|$;
        rotate $(t,\mathcal{P},\Delta)$ so that $k$ becomes $1$
        \label{alg:pollux:cut}
    \State $z^\star \gets \infty$
    \ForAll{$q_1 \in \mathcal{P}_1$}
        \State $D_1(q_1) \gets C(q_1)\Delta_1$;
            $D_1(q) \gets \infty$ for $q \neq q_1$
            \label{alg:pollux:initialize}
        \For{$j = 2,\ldots,N$}
            \State $g \gets \arg\min_{q' \in \mathcal{P}_{j-1}} D_{j-1}(q')$
                \label{alg:pollux:global}
            \ForAll{$q \in \mathcal{P}_j$}
                \State $f \gets \arg\min_{q' \in \mathcal{P}_{j-1},\, R(q',q)=0} D_{j-1}(q')$
                    \label{alg:pollux:free}
                \State $D_j(q) \gets C(q)\Delta_j + \min\{D_{j-1}(f),\, D_{j-1}(g)+\lambda\}$;
                    $\operatorname{parent}_j(q) \gets f$ or $g$ accordingly
                    \label{alg:pollux:update}
            \EndFor
        \EndFor
        \State $q_N \gets \arg\min_{q\in\mathcal{P}_N} D_N(q)+\lambda R(q,q_1)$;
            $z \gets D_N(q_N)+\lambda R(q_N,q_1)$
            \label{alg:pollux:close}
        \State \textbf{if} $z < z^\star$ \textbf{then}
            $z^\star \gets z$;
            $\Gamma^\star \gets \Call{Backtrack}{\operatorname{parent},q_N,q_1}$
            \label{alg:pollux:backtrack}
    \EndFor
    \State \Return $\Call{UndoRotation}{\Gamma^\star,k}$
\end{algorithmic}
\end{algorithm}

\section{Evaluation}
\label{sec:evaluation}

% This section presents a series of experiments as means to evaluate Clockwork. Section~\ref{sec:eval:setup} details the carried-out experiments, experimental frameworks, and evaluation objectives, while Sections \ref{sec:eval:slo}--\ref{sec:eval:sensitivity} present the results.

\subsection{Experimental Setup}
\label{sec:eval:setup}

% We implement Clockwork in Python and evaluate it through trace-driven simulation. We ask whether Castor improves SLO feasibility and reduces provisioning cost relative to single-plane and heuristic baselines, whether Pollux reduces total scheduling cost and reprovisioning while preserving SLO attainment, and how these results vary across deployment conditions. Castor and Pollux execute directly in the prototype. We measure component execution latencies on representative hardware and use them as execution profiles. Orbital traces provide satellite positions, visibility, and ISL propagation latency at each snapshot. The evaluator combines these profiles with payload transmission time to determine SLO attainment.

We implement Clockwork in Python and evaluate it through trace-driven simulation\footnote{The codebase and wiki with complete system model and additional experiments are available at \url{https://github.com/polaris-slo-cloud/Clockwork-LEO}}. We evaluate whether Castor improves SLO feasibility and reduces provisioning cost relative to single-plane and heuristic baselines, whether Pollux reduces scheduling cost and reprovisioning while preserving SLO attainment, and how these results vary across deployment conditions. Castor and Pollux execute as part of the prototype over orbital traces that provide satellite positions, visibility, and ISL propagation latency at each snapshot. We measure component execution latencies on representative hardware and use them as execution profiles. At each snapshot, we compute the end-to-end latency of each provisioning state using \Cref{eq:end-to-end-latency} and determine whether it satisfies the latency SLO.

\begin{table}[h]
\caption{Default experimental configuration. Shell dimensions give planes $\times$ satellites per plane.}
\label{tab:setup}
\centering
\scriptsize
\setlength{\tabcolsep}{3pt}
\renewcommand{\arraystretch}{1.04}

\begin{tabularx}{\columnwidth}{
    @{}p{0.28\columnwidth}>{\raggedright\arraybackslash}X@{}
}
\toprule
\textbf{Parameter} & \textbf{Default value} \\
\midrule

\multicolumn{2}{@{}l}{\textbf{Orbital substrate}} \\
Star shell
& $12\times16$, 561\,km, $87^\circ$; 168 CPU, 20 small, 4 large acc. \\
Delta shell
& $12\times16$, 888\,km, $53^\circ$; 179 CPU, 12 small, 1 large acc. \\
Repeat cycle
& One sidereal day ($86\,164$\,s): 15 Star, 14 Delta revolutions \\
Network
& $+$Grid ISLs 100\,Gbps, downlink 1\,Gbps, min. elevation $25^\circ$ \\
Cost rates
& CPU / small / large 1 / 4 / 10, active plane 5 units per hour \\
\midrule
\multicolumn{2}{@{}l}{\textbf{Request and planning}} \\
Reference request
& $48^\circ$N $16^\circ$E, latency SLO $300$\,ms \\
Castor
& $K=2$ nearest CPU per shell and per plane, margin $\eta=5$ units per hour \\
Pollux
& Reprovisioning cost $\lambda=5$ units \\
Snapshots
& Equally spaced; Castor: 8 snapshots $\times$ 10 seeded hardware assignments; Pollux: 1-min grid, 1,440 $\times$ 3 \\

\bottomrule
\end{tabularx}
\end{table}

\subsubsection{Provider Infrastructure}

\Cref{tab:setup} summarizes the default configuration. We construct synthetic Walker Star and Delta shells~\cite{walker1970some} and generate orbital traces using circular two body motion with Earth rotation. We omit drag and J2 perturbations, an orbital simplification also adopted by Stardust~\cite{pusztai2025stardust}. The selected altitudes produce a common day repeat cycle~\cite{luo2017revisit}.

We represent the CPU, small and large accelerator classes using execution profiles measured on an AMD Ryzen~9~7950X, an NVIDIA Jetson Orin Nano Super, and an NVIDIA RTX~4090, respectively. These commercial off the shelf platforms represent the hardware classes in our model. Their use is motivated by the adoption and proposed deployment of commercial computing hardware in orbit~\cite{denby2020orbital,unibap2026,starcloud2026}.

Accelerators are concentrated in selected planes, so serving the AoI can require combining access and compute resources across planes. In Star, 3 planes each carry 4 small accelerators, while 2 planes each carry 4 small and 2 large accelerators. In Delta, 3 planes each carry 3 small accelerators, while one plane carries 3 small and one large accelerator. CPUs occupy all remaining positions. Each seeded hardware assignment randomly permutes plane configurations within each shell and hardware locations within each plane, preserving each shell's hardware counts and plane configurations. We use normalized provisioning cost rates because commercial prices for orbital compute resources are unavailable.

\subsubsection{Workload and Execution Profiles}
\label{sec:eval:workload}

We evaluate a compound AI workflow that detects a potential wildfire, classifies the detected region, and generates a short report. \Cref{tab:workflow-profiles} lists its models, capacity demands, output sizes, and execution profiles. The input is an 8\,MB image, and each satellite has normalized capacity one.

We report median execution latency at batch size one.
Detector and classifier measurements include preprocessing, inference, and postprocessing. The language model receives a prompt of 85 tokens and generates up to 32 tokens. The small accelerator LLM latency is derived from measured decoding throughput after an execution that terminated after 25 tokens. The CPU configuration does not support the LLM.

\begin{table}[t]
\caption{Compound AI workflow and execution profiles}
\label{tab:workflow-profiles}
\centering
\scriptsize
\setlength{\tabcolsep}{2pt}
\renewcommand{\arraystretch}{1.05}

\begin{tabularx}{\columnwidth}{
  @{}lXccrrr@{}}
\toprule
& & & \textbf{Output}
& \multicolumn{3}{c}{\textbf{Execution latency (ms)}} \\
\cmidrule(lr){5-7}
\textbf{Component}
& \textbf{Model}
& \textbf{Capacity}
& \textbf{(MB)}
& \textbf{CPU}
& \textbf{Small}
& \textbf{Large} \\
& & \textbf{share} & &
& \textbf{accel.}
& \textbf{accel.} \\
\midrule

Detector
& YOLOv8m
& 0.50
& 2.80
& 90.1
& 96.6
& 9.0 \\

Classifier
& ResNet-50
& 0.25
& 0.04
& 18.7
& 24.6
& 2.1 \\

LLM
& Qwen2.5-0.5B
& 0.75
& 0.01
& n/a
& 900
& 117 \\

\bottomrule
\end{tabularx}

\end{table}

\subsubsection{Baselines}
  \label{sec:eval:baselines}

% In the absence of existing approaches that address the same problem as Clockwork, we construct representative baselines that capture fundamental trade-offs of scheduling compound AI workflows in space. First, we compare Castor with three provisioning baselines.
% \textsc{Best Single Plane} (BSP) finds the least cost valid state confined to one plane, isolating the benefit of multi plane composition. \textsc{Greedy Composition} (GC) places components sequentially on the compatible satellite with the lowest additional cost while reserving sufficient latency for the remaining workflow, isolating the benefit of Castor's joint placement search. \textsc{Latency First} (LF) searches the same candidates as Castor but minimizes latency, isolating the benefit of cost aware selection.

We compare Castor with three provisioning baselines. \textsc{Best Single Plane} (BSP) finds the least cost valid state confined to one plane, isolating the benefit of multi plane composition. \textsc{HyperDrive}~\cite{pusztai2024hyperdrive} adapts the placement policy of \textsc{HyperDrive} to our snapshot model, using Castor's candidate set and execution-aware per-edge latency budgets. It places components sequentially, ranking candidates by network latency, isolating the benefit of Castor's joint placement search against an external baseline. \textsc{Latency First} (LF) searches the same candidates as Castor but minimizes latency, isolating the benefit of cost aware selection.

We compare Pollux with three scheduling baselines. \textsc{Per Snapshot Minimum} (PSM) selects the least cost valid state independently at each snapshot, measuring the cost of ignoring future transitions. \textsc{Reactive} retains the current state until it violates the SLO, representing adaptation without lookahead. \textsc{Rolling Horizon} (RH) optimizes over the next \(H\) minutes and applies its first decision, measuring how much lookahead is needed to approach cycle wide planning. All methods use the same Castor candidate states and cost model.

\subsection{Per-Snapshot Provisioning under Latency SLOs}
\label{sec:eval:slo}

We first evaluate whether Castor's joint search increases SLO feasibility and reduces provisioning cost. We vary the latency SLO from 200 to 2000\,ms using the default fleet and reference AoI. For each SLO, we evaluate eight evenly spaced planning epochs for each of ten seeded hardware assignments, giving 80 instances. We compare Castor with the \textsc{BSP}, \textsc{HyperDrive} and \textsc{LF}.

\begin{figure}[h]
      \centering
      \includegraphics[width=0.8\columnwidth]{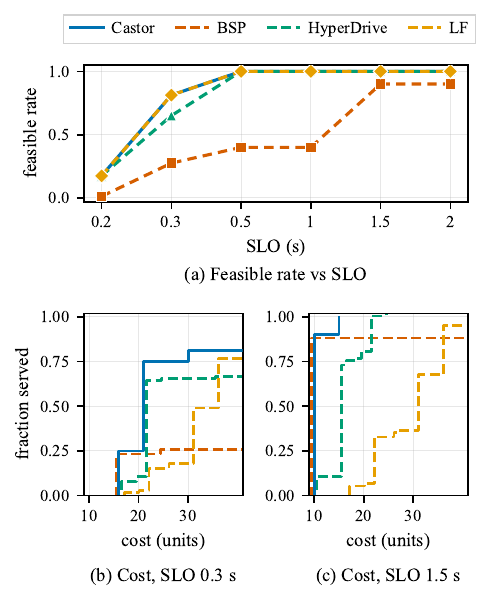}
      \caption{Effect of the latency SLO on snapshot provisioning. (a) Feasibility vs various SLOs. (b) and (c) Fraction of instances served at or below each provisioning cost at the 300\,ms and 1500\,ms SLO.}
      \label{fig:slo-sweep}
  \end{figure}

\Cref{fig:slo-sweep}~(a) shows how the latency SLO affects the fraction of instances that each method can serve. Below 200,ms, all methods fail to find a valid state. At 200\,ms, Castor serves 18\% of the instances, compared with 8\% for \textsc{HyperDrive} and 1\% for \textsc{BSP}. At the default 300\,ms SLO, Castor serves 81\%, compared with 65\% for \textsc{HyperDrive} and 28\% for \textsc{BSP}. The gap over \textsc{BSP} shows that meeting a tight SLO often requires resources from multiple planes. The gap over \textsc{HyperDrive} shows that selecting these resources jointly finds valid states that sequential placement misses. The advantage narrows as the SLO loosens. At 500\,ms, \textsc{HyperDrive} and Castor serve every instance, while \textsc{BSP} serves only 40\%. Thus, HyperDrive's sequential placement becomes sufficient for feasibility, but a single plane often remains insufficient. \textsc{BSP} reaches 90\% at 1500\,ms, when the small accelerator can execute the LLM within the latency SLO.

\Cref{fig:slo-sweep}~(b)--(c) show the fraction of instances served by plans at or below each provisioning cost. At 300\,ms, Castor serves 75\% of the instances within 21 units, compared with 65\% for \textsc{HyperDrive}, 25\% for \textsc{BSP}, and 16\% for \textsc{LF}. \textsc{LF} requires 40 units to reach Castor's maximum feasibility of 81\%. This occurs because only the large accelerators can execute the LLM under the tight SLO, which often requires Castor to combine resources across planes. At 1500\,ms, Castor, \textsc{HyperDrive}, and \textsc{LF} serve every instance, but Castor reaches this value by 16 unit cost while \textsc{LF} requires 40. \emph{Thus, Castor's joint composition improves feasibility under the tight SLO, while cost-aware selection avoids unnecessary accelerator cost on loose SLOs.}

\subsection{Cycle-Wide Scheduling}
\label{sec:eval:pollux}

We evaluate whether planning across snapshots reduces total scheduling cost while preserving SLO attainment. Using the default configuration, we compare Pollux with the scheduling baselines over one repeat cycle across three seeded hardware assignments. All methods use the same candidate pool for each assignment and reprovisioning cost $\lambda=5$, equal to the plane activation cost. We also examine cumulative cost during the first two hours of one assignment to illustrate how scheduling decisions affect cost over time.

\begin{figure}[h]
    \centering
    \includegraphics[width=0.8\columnwidth]{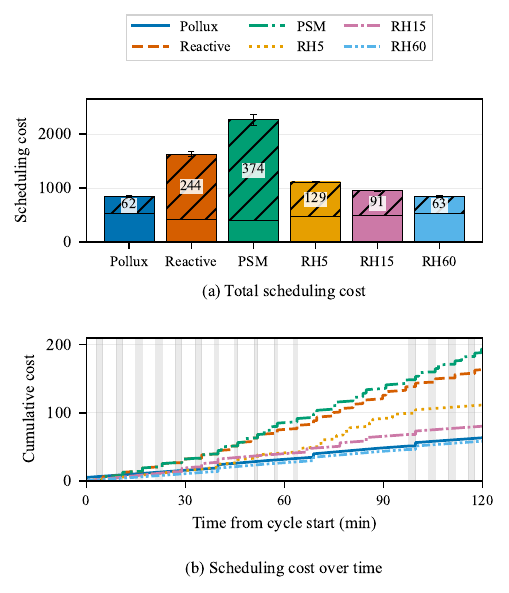}
    \caption{
     (a) Mean total cost across three seeded hardware configurations, split into provisioning (solid) and reprovisioning (hatched) cost.
    (b) Cumulative cost during the first two hours for one seeded hardware configuration. Gray bands mark periods with no SLO feasible candidate plan.
    }
    \label{fig:pollux}
\end{figure}

\Cref{fig:pollux}~(a) shows that Pollux reduces total scheduling cost by 49\% relative to \textsc{Reactive} and 63\% relative to \textsc{PSM}, while preserving the same SLO attainment. Its mean unit cost is 837, compared with 1,630 and 2,269, respectively. Pollux makes 62 reprovisionings, compared with 244 and 374, offsetting its higher provisioning cost. All methods meet the SLO at approximately 84\% of snapshots, with no feasible candidate at the remaining snapshots. Using the same dynamic program with \textsc{RH60} also approaches Pollux's cost, indicating that this amount of lookahead captures most of the scheduling benefit in the evaluated configurations. Pollux achieves these savings with one schedule computed for the repeat cycle, showing the benefit of jointly planning provisioning decisions
and their changes over time.

\Cref{fig:pollux}~(b) shows how the cost difference develops over time. Provisioning costs accumulate between changes, while charged reprovisionings produce jumps. \textsc{Reactive} and \textsc{PSM} repeatedly release resources during short service gaps and incur reprovisioning costs when service resumes. Pollux retains resources through these gaps, making four charged reprovisionings in the displayed interval, compared with 26 and 32, respectively. Longer rolling horizons also reduce cumulative cost, with \textsc{RH60} tracking Pollux closely. \emph{The trace illustrates that holding resources during a short service gap costs less than releasing and reprovisioning them.}

\subsection{Sensitivity to Deployment Conditions}
\label{sec:eval:sensitivity}

We next test whether the preceding results depend on the reference deployment. We vary the availability and distribution of accelerator hardware, the AoI latitude and available orbital shells, and whether the shell geometry repeats over the planning cycle. These experiments test the generality of Castor's provisioning results and identify the orbital conditions under which a Pollux schedule must be recomputed.

\begin{figure}[h]
  \centering
  \includegraphics[width=0.8\columnwidth]{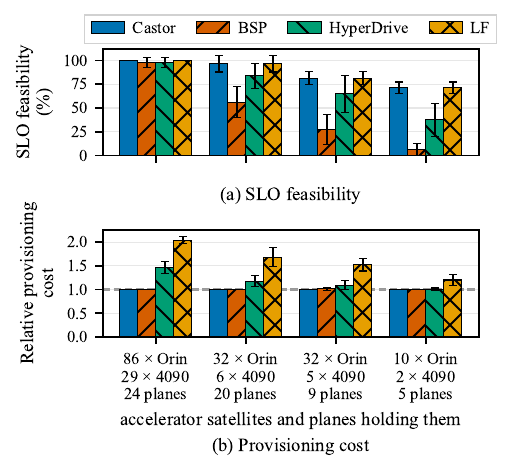}
  \caption{Effect of accelerator availability on placement.}
  \label{fig:hardware-sensitivity}
\end{figure}

\paragraph{Hardware availability}

To test whether Castor's advantage depends on the reference hardware distribution, we compare it with the provisioning baselines under four hardware configurations that vary how many accelerator satellites the constellation holds and how many planes hold them. The first spreads 86 Orin and 29 RTX 4090 satellites over all 24 planes. The second spreads 32 Orin and 6 RTX 4090 satellites over 20 planes. The reference configuration concentrates 32 Orin and 5 RTX 4090 satellites in 9 planes, and the fourth concentrates 10 Orin and 2 RTX 4090 satellites in 5 planes. CPUs occupy all remaining satellites, and all other parameters remain unchanged.

\Cref{fig:hardware-sensitivity}~(a) shows that Castor's advantage increases as accelerators become scarcer and more concentrated. All methods serve at least 93\% of instances in the accelerator-rich fleet. With the fleet-wide control, Castor serves 96\% of instances, compared with 78\% for \textsc{HyperDrive} and 56\% for \textsc{BSP}. These rates fall to 81\%, 65\%, and 28\%, respectively, in the reference fleet, and to 71\%, 38\%, and 6\% in the scarce fleet. \Cref{fig:hardware-sensitivity}~(b) compares provisioning cost at epochs served by both Castor and each baseline. \textsc{LF} matches Castor's feasibility but costs between 1.2 and 2.0 times as much. Castor is never more expensive than a baseline at any jointly served epoch. \emph{These results show that plane composition and joint search become increasingly important as accelerator access becomes more limited.}

% \paragraph{AoI and shell selection.}
% We vary the AoI latitude from \(0^\circ\) to \(75^\circ\) using identical hardware distributions in both shells, which isolates the effect of orbital geometry. Across these latitudes, allowing Castor to select either shell serves 93--100\% of instances. Star alone serves 75\% below \(60^\circ\) and reaches 100\% at \(75^\circ\), while Delta serves 83--95\% up to \(60^\circ\) but cannot serve \(75^\circ\). Provisioning cost remains between 16 and 21 units. The shells therefore provide complementary geographic availability, while Castor selects one shell for each provisioning state.

% \begin{figure}[h]
%   \centering
%   \includegraphics[width=0.8\columnwidth]
%       {Figures/Evaluation/pBC_geography.pdf}
%   \caption{Effect of AoI latitude and shell availability. Lines show
%   means over five hardware assignments, and error bars show one standard
%   deviation.}
%   \label{fig:aoi-sensitivity}
% \end{figure}

\paragraph{Orbital repeatability}

\begin{table}[t]
\caption{Effect of orbital repeatability on the Pollux schedule. Brackets give the min and max replay result across seeds.}
\label{tab:orbital-repeatability}
\centering
\scriptsize
\setlength{\tabcolsep}{3pt}
\renewcommand{\arraystretch}{1.05}
\begin{tabular}{@{}lrrrr@{}}
\toprule
& \multicolumn{3}{c}{\textbf{Day-two SLO attainment (\%)}} &
\textbf{Reprov.} \\
\cmidrule(lr){2-4}
\textbf{Shell altitudes (km)}
& \textbf{Ceiling}
& \textbf{Replay}
& \textbf{Replan}
& \\
\midrule
Repeat, 561/888
& 84.3 & 84.3 & 84.3 & 62 \\
Non-repeat, 550/800
& 83.7 & 51.4 [49--53] & 83.7 & 61 \\
Non-repeat, 600/1000
& 85.8 & 47.1 [42--50] & 85.8 & 59 \\
\bottomrule
\end{tabular}
\end{table}

Repeat-track shells allow Pollux to compute one cyclic provisioning schedule and reuse it on subsequent days. We test whether schedule reuse depends on this property by moving the shells to non-repeat altitudes of 550/800\,km and 600/1000\,km (\Cref{tab:orbital-repeatability}). On the reference shells, replaying the first-day schedule attains 84.3\% SLO compliance on the second day, equal to the maximum achievable for that day's geometry. On the non-repeat shells, replay attains only 51.4\% and 47.1\% because the visibility intervals shift between days. Recomputing from the predicted second-day geometry restores SLO compliance to the respective maxima of 83.7\% and 85.8\%, with 61 and 59 reprovisionings compared with 62 on the reference shells. This shows that the loss under replay is caused by changing orbital geometry rather than a limitation of Pollux. \emph{Repeat-track shells therefore permit schedule reuse, whereas other shells require Pollux to recompute the schedule as their geometry changes.}
\section{Discussion}

Clockwork deliberately targets stateless compound AI workflows, whose explicit components, payloads, and per-hardware execution profiles make placement and communication costs tractable to model. Stateful applications would additionally require state migration and consistency during reprovisioning, while less structured applications may be harder to characterize with sufficiently accurate execution profiles. Clockwork exploits predictable orbital motion but assumes that resource and network availability follow the planned constellation state. Satellite or link failures, temporary communication loss, and power or thermal constraints could therefore invalidate planned provisioning decisions. Our evaluation further assumes uncongested communication, static execution profiles, normalized provisioning costs, and a fixed reprovisioning cost. These abstractions support controlled comparison of provisioning strategies but do not predict operational deployment costs or transition times. Finally, Clockwork evaluates feasibility on a one-minute planning grid because provisioning changes are driven by visibility intervals lasting several minutes. This provides multiple decision points within each interval and is sufficient for the provisioning and scheduling comparisons made here, though it does not establish continuous SLO satisfaction between snapshots.

\section{Related Work}
\label{sec:related}

Prior work relevant to Clockwork spans two areas: (1) serving compound AI workflows on fixed infrastructure and (2) placing workloads on LEO constellations.

\subsection{Serving Compound AI Workflows}

Deploying a compound AI workflow means choosing, for each of its components, a model variant and the hardware it runs on, under latency, cost, and accuracy objectives~\cite{gravara2026design}. Serving systems make this choice on fixed infrastructure. IPA and Loki select variants and scale hardware so that a pipeline meets its latency SLO at low cost~\cite{ghafouri2024ipa,ahmad2024loki}; Atlas places variants on heterogeneous hardware per stage to maximize predicted accuracy under SLOs~\cite{gravara2026atlas}; Compass switches between precomputed configurations as load changes~\cite{gravara2026compass}; and Murakkab maps agentic workflows to models and hardware and reconfigures them at runtime~\cite{chaudhry2026murakkab}. Clockwork inherits their view of a workflow as a chain of components with per-hardware execution profiles, and their principle that placement is decided by an SLO and a cost.

These systems assume that every node is reachable for the whole deployment, so a placement, once chosen, stays feasible until the load changes, and only the load triggers adaptation. On a constellation, however, the nodes leave the view of the AoI within minutes, so feasibility changes with geometry, and the placement decision recurs over time with a cost for every change. These systems therefore cannot be applied directly to a constellation, and Clockwork builds the temporal dimension they lack on top of the per-component configuration they solve.

\subsection{Workload Placement on LEO Constellations}

Satellite computing systems place tasks, microservices, and pipelines on satellites and across the edge-cloud-space continuum, and handle orbital motion through handovers and migration~\cite{li2025orbitchain,huang2023satedge,lin2025dependency,bhosale2024krios,zheng2025sdn}. Service-chain placement over time-expanded LEO graphs plans across successive snapshots and limits migrations of network functions~\cite{mahyoub2026scalable,liu2026vmr,liu2025switching}, building on the predictability of orbital motion that contact-plan routing established for networking~\cite{wood2001effects}. Clockwork inherits this predictability and the handover of access satellites within a fixed placement. HyperDrive~\cite{pusztai2024hyperdrive} is closest to our per-snapshot problem: it schedules serverless functions across edge, cloud, and satellite nodes under latency objectives, ranking candidates by network proximity. Its placement policy operates on a single topology snapshot and does not model provisioning cost or reprovisioning over time.

These systems decide where a workload runs on satellites the request can already use, and where they plan over time, they do so for uniform functions on individual satellites. Clockwork decides which planes to activate for an AoI and where to place components with different hardware classes, and weighs the provisioning cost of holding a state against the reprovisioning cost of changing it over the cycle.

\begin{table}[t]
\caption{Related work by group. \checkmark{} all systems in the group, (\checkmark) some, -- none. Columns: model selection; hardware heterogeneity; placement feasibility with orbital motion; placements planned ahead over snapshots; provisioning cost in the objective.}
\label{tab:related}
\centering
\scriptsize
\setlength{\tabcolsep}{2pt}
\renewcommand{\arraystretch}{1.05}
\begin{tabular}{@{}p{3.95cm}ccccc@{}}
\toprule
& \textbf{Model} & \textbf{HW} & \textbf{Orbital} & \textbf{Plans} & \textbf{Prov.} \\
& \textbf{selection} & \textbf{classes} & \textbf{motion} & \textbf{ahead} & \textbf{cost} \\
\midrule
\raggedright Compound AI serving~\cite{ghafouri2024ipa,ahmad2024loki,gravara2026atlas,gravara2026compass,chaudhry2026murakkab}
& \checkmark & (\checkmark) & -- & -- & \checkmark \\
\raggedright LEO placement~\cite{bhosale2024krios,lin2025dependency,zheng2025sdn,li2025orbitchain}
& -- & (\checkmark) & \checkmark & (\checkmark) & (\checkmark) \\
\raggedright Time-expanded chains~\cite{mahyoub2026scalable,liu2026vmr,liu2025switching}
& -- & -- & \checkmark & \checkmark & \checkmark \\
\raggedright HyperDrive~\cite{pusztai2024hyperdrive}
& -- & \checkmark & (\checkmark) & -- & -- \\
\textbf{Clockwork}
& -- & \checkmark & \checkmark & \checkmark & \checkmark \\
\bottomrule
\end{tabular}
\end{table}

% Sections/conclusion.tex

\section{Conclusion}
\label{sec:conclusion}

Serving compound AI workflows from LEO constellations requires jointly deciding which planes to activate, where to place workflow components, and when these decisions should change as orbital geometry evolves. We presented Clockwork, which uses predicted orbital geometry to plan these provisioning decisions over time. Castor searches each snapshot for low cost feasible provisioning states, while Pollux connects these states across snapshots and
accounts for both provisioning and reprovisioning cost. In our evaluation, Castor serves 81\% of instances at a 300\,ms SLO, while Pollux reduces mean total scheduling cost by up to 63\% with equal snapshot SLO attainment. These results show that provisioning on a constellation is a decision in time as well as in space, and that predictable orbital motion enables these decisions to be planned in advance.

Future work includes multi tenant scheduling for shared orbital resources, support for stateful workflows that require state migration or synchronization during reprovisioning and incorporation of satellite and link failures, resource contention, and changing power or thermal availability to improve robustness in operational space data centers.

% --- Acknowledgment (comment out for blind submission) ---
\section*{Acknowledgment}
This work was partly funded by the European Union under the Horizon Europe programme through the SNS JU (Grant Agreement No. 101192912, NexaSphere). Views expressed are those of the authors and do not necessarily reflect those of the EU or the SNS JU.

% --- References ---
\bibliographystyle{IEEEtran}
\bibliography{references}

% Balance the two columns on the final page:
\balance

\end{document}